\documentclass[twocolumn,useAMS,usenatbib]{mn2e}
\usepackage{psfig}

\title[Distance and progenitor of SN~2005cs]{Distance estimate and progenitor characteristics 
of SN~2005cs in M51}
\author[K. Tak\'ats and J. Vink\'o]{K. Tak\'ats$^1$\thanks{e-mail: takats.katalin@stud.u-szeged.hu}
 and J. Vink\'o$^1$\thanks{e-mail: vinko@physx.u-szeged.hu}\\
 $^1$Department of Optics and Quantum Electronics, University of Szeged, D\'om t\'er 9., Szeged, 
 Hungary}
\begin{document}

\date{Accepted; Received; in original form}

\pagerange{\pageref{firstpage}--\pageref{lastpage}} \pubyear{2006}

\maketitle

\label{firstpage}

\begin{abstract}
Distance to the Whirlpool Galaxy (M51, NGC~5194) is estimated using published photometry
and spectroscopy of the Type II-P supernova SN 2005cs. Both the Expanding
Photosphere Method (EPM) and the Standard Candle Method (SCM), suitable
for SNe II-P, were applied. The average distance ($7.1 \pm 1.2$ Mpc) is
in good agreement with earlier SBF- and PNLF-based distances, but slightly longer
than the distance obtained by \citet{bar} for SN 1994I via the Spectral Fitting
Expanding Atmosphere Method (SEAM). 
Since SN 2005cs exhibited low expansion velocity
during the plateau phase, similarly to SN 1999br, the constants of SCM were
re-calibrated including the data of SN 2005cs as well. The new relation is
better constrained in the low velocity regime ($v_{ph}(50) \sim 1500 - 2000$ 
km s$^{-1}$), that may result in better distance estimates for such SNe. 
The physical parameters of SN~2005cs and its progenitor is re-evaluated
based on the updated distance. All the available data support the low-mass 
($\sim 9 ~M_\odot$) progenitor scenario proposed previously by its direct detection
with the $Hubble ~Space ~Telescope$ \citep{maund,li}.
\end{abstract}


\section{Introduction}
The Type II-P SN~2005cs in M51 was discovered by \cite{kloehr} on June 28.9 2005. 
The first spectroscopic data \citep{modjaz} indicated that this SN was caught 
at very early phase, shortly after explosion. \citet{past} presented high-quality
$UBVRI$ photometry and optical spectroscopy obtained during the first month of
the plateau phase. From their data supplemented by amateur observations they
could determine a tight constraint on the explosion time. They derived 
JD $2453549 \pm 1$ (June 27.5 UT, 2005), which is adopted in this paper, 
and will be used for the distance determination later. They also collected
the available information for the reddening of SN 2005cs, and found 
$E(B-V) = 0.11 \pm 0.04$ mag, which is consistent with the blue colour
of SN 2005cs at the early phase (see Pastorello et al., 2006 for discussion).
\citet{tsvet} published additional photometry extending into the nebular
phase, from which they estimated the explosion time very close to that of
\citet{past} (within $0.2$ day) and a nickel mass $M_{Ni} \sim 0.018 ~ M_\odot$.

The progenitor of this SN was studied by \cite{li} and \cite{maund} 
using archival $HST$ images. Both teams have detected the possible
progenitor, but only in the $I$ band, which led to the conclusion that
the progenitor was probably a red giant. From the $I$ band flux and the
flux upper limits in the other bands the mass of the progenitor turned
out to be relatively small: $M_{ZAMS}=7-9 ~M_\odot$ \citep{li} 
and $M_{ZAMS}=7-12 ~M_\odot$ \citep{maund}. The lower limit of these mass
estimates are close to the theoretical limit of core collapse 
\citep{hilleb}. 

In this paper we present a new distance estimate to M51 based on the
published data of SN 2005cs. The application of the distance measurement
methods are in Section 2. In Section 3 we compare the SN-based distance
to other recent distance estimates and derive an average value that is
consistent with the available information. Finally, we update the
physical parameters of SN~2005cs based on the new distance.

\section{Distance measurement}
In this section we apply the Expanding Photosphere Method (EPM) and
the Standard Candle Method (SCM) for estimating the distance to SN~2005cs.

\subsection{Expanding Photosphere Method}

\begin{table*}
\begin{center}
\caption{Quantities determined in EPM. The columns contain the followings: observational 
 epoch (JD$-$2450000), bolometric flux (in erg s$^{-1}$ cm$^{-2}$), temperature (in K),
 velocity at the photosphere (in km s$^{-1}$),
 angular size (in $10^8$ km Mpc$^{-1}$) and $\theta/v_{ph}$ (in day Mpc$^{-1}$). 
 The uncertainties are in parentheses.}
\begin{tabular}{cccccc}
\hline
 $t$ &  $f_{bol}$  & $T_{eff}$ & $v_{ph}$ & $\theta$ & $\theta/v_{ph}$ \\
 (JD-2540000) & (erg s$^{-1}$ cm$^{-2}$) & (K) & km s$^{-1}$ &($10^8$ km/Mpc) & (day/Mpc) \\  
\hline
3552.36&5.11$\cdot 10^{-11}$&35864 (4207) & 6370(300) &0.4158 & 0.0752 \\
3553.35&5.28$\cdot 10^{-11}$ &30183 (1220) & 5900(300) &0.6093 & 0.1195 \\
3554.46&5.07$\cdot 10^{-11}$ &23624 (180)  & 5550(300) &1.0042 & 0.2079 \\
3557.42&4.94$\cdot 10^{-11}$ &20058 (243)  & 5050(300) &1.3997 & 0.3146 \\
3557.84&4.68$\cdot 10^{-11}$ &16836 (2092) & 4900(300) &1.9601 & 0.4630 \\
3559.40&4.62$\cdot 10^{-11}$ &15038 (1633) & 4560(300) &2.4489 & 0.6095 \\
3563.38&4.46$\cdot 10^{-11}$ &10235 (1417) & 4030(250) &4.9404 & 1.4206 \\
3563.42&4.35$\cdot 10^{-11}$ & 9385 (1730) & 4010(250) &5.6194 & 1.6199 \\
3565.38&4.11$\cdot 10^{-11}$ & 9244 (1468) & 3725(250) &5.5881 & 1.7480 \\
3566.36&4.04$\cdot 10^{-11}$ & 8807 (1195) & 3600(250) &5.9649 & 1.9177 \\
3566.40&3.98$\cdot 10^{-11}$ & 8999 (1533) & 3600(250) &5.7287 & 1.8469 \\
3569.42&3.81$\cdot 10^{-11}$ & 7575 (1568) & 3330(250) &7.0943 & 2.4510 \\
3571.40&3.72$\cdot 10^{-11}$ & 7495 (1429) & 3180(250) &7.1004 & 2.5843 \\
3572.40&3.68$\cdot 10^{-11}$ & 7341 (1512) & 3090(250) &7.2426 & 2.7040 \\
3577.40&3.72$\cdot 10^{-11}$ & 7049 (1479) & 2730(250) &7.6167 & 3.3267 \\
3579.40&3.70$\cdot 10^{-11}$ & 6519 (1415) & 2580(250) &8.2323 & 3.8112 \\
3583.39&3.66$\cdot 10^{-11}$ & 6023 (1537) & 2260(250) &8.7738 & 4.4152 \\
3583.47&3.64$\cdot 10^{-11}$ & 6051 (1630) & 2260(250) &8.7205 & 4.4858 \\
\hline	        		
\end{tabular}   		
\label{data}
\end{center}
\end{table*}

The Expanding Photosphere Method \citep{kir} derives distance of an optically
thick, homologously expanding SN atmosphere, radiating as a diluted blackbody. 
These assumptions are expected to be close to the real physical situation
in a dense atmosphere of a Type II-P SN at the early plateau phase, when
the ejecta is almost fully ionized, and electron scattering dominates the
true absorption \citep{dess}. Since SN 2005cs was observed in the first
30 days of the plateau phase, this is a promising object for the application
of EPM.

We have applied the ,,bolometric'' version of EPM described by \citet{vinko}.
The $BVRI$ photometric data from \citet{past} were 
dereddened using $E(B-V)=0.11$, then transformed into fluxes using 
the calibration given by \citet{ham2}:
\begin{equation}
 f_\lambda = {h \cdot c \over \lambda \cdot W_\lambda} \cdot 10^{(m_0-m)/2.5}   
\end{equation}
where $W_\lambda$ is the $FWHM$ of the given filter, $m$ 
is the dereddened magnitude, $m_0$ is the zero-point of the magnitude scale.

From these quasi-monochromatic fluxes, the bolometric flux $f_{bol}$ was 
estimated for each epochs by numerically integrating $f_\lambda$
using the effective wavelengths and $FWHM$ values of the $BVRI$ filters. 
The missing fluxes in lower and greater wavelengths were extrapolated lineary 
from the $B$- and $I$-band fluxes, assuming zero flux at 
3400 \AA~ and 23000 \AA~ \citep{vinko}. 

The angular radius was derived from the bolometric light curve as
\begin{equation}
 \theta = \sqrt{f_{bol} \over \zeta^2(T) ~\sigma ~T_{eff}^4}
\label{1}
\end{equation} 
where $f_{bol}$ is the observed bolometric flux, $T_{eff}$ is the effective 
temperature and $\zeta(T)$ is the dilution factor. 

The effective temperatures were estimated by fitting a blackbody to the fluxes 
in $B$, $V$ and $I$ bands. The $R$ band was omitted because of the presence 
of the $H\alpha$ emission line. 
\citet{vinko} has recently demonstrated that the total flux of
the blackbody corresponding to temperature $T_{BVI}$ produces a reasonable
distance estimate, although the spectral flux distribution of the SN is, of course,  
not exactly Planckian. Fortunately, the deviation from the blackbody curve
is not so severe at the early phases, when SN 2005cs was observed, but
becomes more and more pronounced after $t ~>~ 20$ days, when the lines of 
ionized metals appear in the spectrum.

The dilution factor $\zeta(T)$ comes from model atmospheres. We have
applied the dilution factors derived recently by \citet{dess}. 
They determined $\zeta$ for the first 30 days of the plateau.
They also argued that EPM should be applied for phases earlier
than 30 days, because of the failure of the blackbody assumption
at later phases, when metallic lines dominate the photospheric 
spectrum. Note that their model atmospheres had $T_{eff} < 20000$ K,
thus, the dilution factors are valid only in the range of 
$4000 ~<~ T_{eff} ~<~ 20000$ K (see Fig.~1 of \citet{dess}).

The dilution factors of \citet{dess}
are systematicaly higher than those of \citet{east}. This has the
consequence that the distances computed from the \citet{dess} dilution
factors are systematically longer than those from the \citet{east}
dilution factors. It is not clear why these two sets of model atmospheres
give different dilution factors, but \citet{dess2} provided a convincing
evidence that their data result in a distance to SN 1999em, which is in
good agreement with the Cepheid-based distance to the host galaxy, while
the calculation based on the earlier set of dilution factors give a
distance that differs by 50 \% \citep{leonard3}. 

We have calculated the distance by applying the linear equation
\begin{equation}
 t = t_0 + D {\theta \over v_{ph}}
\label{2} 
\end{equation}  
where $t_0$ is the moment of the explosion, $v_{ph}$ is the photospheric 
expansion velocity at epoch $t$ and $D$ is the distance.

The velocity data were selected from \cite{past} and were interpolated to the epochs of the 
photometric data. In the first days, when there are no metallic lines in the spectrum, we 
used the velocities inferred from the He~I $\lambda 5876$ line. This line is a good
indicator of the photospheric velocity at the early phases. According to
our parametrized model spectra of Type II-P SNe (Vink\'o, in preparation), the velocities from 
He~I $\lambda 5876$  match the input photospheric velocities of the model spectra
within $\pm 2$ \%. After $JD = 2453557$ the Fe~II $\lambda 5169$
line was used, which is the standard one for computing photospheric velocities in SNe II-P
atmospheres \citep{dess}.

The derived parameters for SN~2005cs are listed in Table \ref{data}. 
Note that the first 6 points were omitted from the fitting, due to
the following reasons. The estimated temperature of the first 4 data
is over 20000 K, outside the temperature range of the dilution 
factors (see above). The next two points have the highest errorbars of
their fitted $T_{eff}$, and the theoretical dilution factors of \citet{dess}
also show large scatter around these temperatures. Moreover, our approach
for computing the bolometric fluxes has been tested to work well only for
$T < 10000$ K \citep{vinko}. The bolometric fluxes are underestimated
above this temperature, due to the very approximative treatment of the
UV flux, which has significant contribution at high temperatures.

\begin{figure}
\begin{center}
\psfig{file=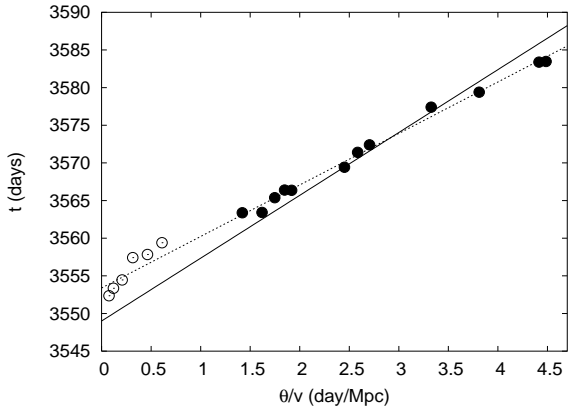,width=8cm}
\end{center}
\caption{Distance determination with EPM. The distance and the explosion epoch 
 are obtained from the fitted line. Continuous line: fixed explosion 
 epoch; dotted line: fitted explosion epoch. The data marked by open circles
 were omitted from the fitting (see text), but their inclusion do not change
 the results significantly.}
\label{D}
\end{figure}

Because the moment of the explosion of SN 2005cs could be determined with 
very small uncertainity ($JD = 2453549 \pm 1$ \citet{past}), at first, we kept
this parameter fixed and derived only the 
distance. This resulted in $D = 8.34 \pm 0.30$ Mpc (Fig. \ref{D}). The $\pm 1$ day 
uncertainity of the explosion epoch changes the distance with $\pm 0.34$ Mpc. 

Secondly, when $t_0$ was treated as a free parameter, the distance decreased 
to $D = 6.84 \pm 0.18$ Mpc. The explosion epoch turned out to be 
$t_0 = 2453553.39 \pm 0.52$ JD. This is 4 days later than the one coming from the
photometric data, and clearly inconsistent with the observations (the discovery date
is 2 days earlier). However, the slightly lower distance describes the observed
data better, than the previous one (see Fig. \ref{D}). 

It is difficult to explain why the two EPM solutions give different distances by
$\sim 20$ \%.
The inclusion of the first 6 data that were disregarded before (see above) does 
not change either solution significantly. In the first case, they are too close
to the explosion epoch which is fixed, thus, the slope of the fitted line
remains unchanged. In the second case, when the explosion epoch is also fitted,
surprisingly, the first 6 points lie nicely close to the line fitted to the other
data. Thus, the varying uncertainties of our bolometric fluxes, although undoubtedly 
present, cannot fully explain the $\sim 20$ \% inconsistency between the EPM solutions
of fixed and variable explosion epoch. 

Another possible reason is the systematic error in the measurement of the 
photospheric velocity. Since we are using published velocities, and the
original spectra are not at our disposal, we cannot quantify the amount
of their potential systematic error better than estimating the errorbars
according to the resolution of the spectra given by \citet{past}. 
Then, as a test, we have systematically increased all 
velocities in Table \ref{data} by $1~\sigma$ and refitted $t_0$ and $D$. 
As a result, the distance increased to $7.5 \pm 0.2$ Mpc, but the explosion
epoch remained the same, $t_0 = 2453353.37 \pm 0.4$ JD. Thus, systematic
underestimate of the velocities may result in the underestimate of the distance,
but it does not solve the problem of the computed explosion epoch that is still 
too late. 

The most problematic part of EPM is, of course, the issue of the dilution
factors that are computed from complex models of SNe atmospheres \citep{dess}. 
Despite the continuous efforts for improving them, they may still contain some sort of
systematic uncertainty, beside the statistical errors that are 
$\sim 5$ \% between $4000 ~<~ T_{BVI} ~<~ 8000$ K, but increase up to 
$\sim 20 - 30$ \% for $T_{BVI} ~>~ 12 000$ K (see Fig. 1 of \citet{dess}).
For comparison, the whole analysis was repeated with the use of the dilution
factors by \citet{east} that are systematically lower at a given temperature 
(see above). As expected, this resulted in significantly lower
distances: $D = 6.37 \pm 0.12$ Mpc (with $t_0 = 2453349$ fixed) and $D = 5.40 \pm 0.13$ Mpc 
(with $t_0 = 2453353.8 \pm 0.4$ fitted). Their average, $5.88$ Mpc is below the other 
M51 distances determined independently (see Sect.~3), similarly to the case of SN~1999em
mentioned above. These results suggest that the \citet{dess} dilution
factors provide a better EPM-distance when applied to data obtained 
during the first month of the photospheric phase, at least in the case of
SN~2005cs. However, regardless of the dilution factors, 
there seems to be a persistent problem with the explosion date, i.e. the best-fitting
EPM solution predicts an explosion epoch which is clearly too late with respect
to the earliest observations. This is a warning sign that the assumptions
of EPM, for example the spherically symmetric ejecta, may be incorrect. 

Nevertheless, we decided to consider the average of the two original EPM-solutions in
estimating the final distance (see Section 3). 
Therefore, $D = 7.59 \pm 1.02$ Mpc has been adopted as the best EPM-distance to 
SN~2005cs.

Note that Eq. 3 was derived by neglecting the radius of the progenitor with
respect to the term $v_{ph} (t-t_0)$, as usual. Taking into account the
progenitor radius $R_0$ in $\theta ~=~ (R_0 ~+~ v_{ph} (t - t_0)) / D$ 
increases the EPM-distance. In the case of SN~2005cs, 
setting $R_0 = 500 ~R_\odot$ results in a 6 \% increase of the distance. 
However, it is shown in Sect. 3 that SN~2005cs probably
had a smaller initial radius of $\sim 180 ~R_\odot$, which has negligible
effect on the distance (about 2 \%).

\subsection{Standard Candle Method}

The Standard Candle Method (SCM) is based on a luminosity-velocity 
relation at 50 days after explosion, approximately in the middle of the plateau phase.
This method was calibrated using 24 SNe \citep{ham1}. More recently, \citet{nugent} refined
the relation by adding two local SNe to the calibrating sample, re-formulated
and extended the method to cosmological distances. 

We tried to apply SCM for SN~2005cs in two ways. First, the
original relation by \citet{ham1} was considered. 
For this purpose, one needs the $V$ and $I$ magnitudes and the expansion 
velocity obtained on the 50th day after explosion. Since the data of \citet{past} 
do not reach this phase, we used the photometry of \cite{tsvet}, which is
in good agreement with the data of \citet{past}. $V(50)=14.69 \pm 0.1$ mag 
and $I(50)=13.96 \pm 0.1$ mag was determined by linear interpolation. 
Unfortunately, there are no published velocity data of SN 2005cs extending
into day $+50$, so we had to estimate this parameter. First, the
velocity curves of SN 2005cs and the other low-velocity Type II-P SN~1999br
\citep{hamphd} was matched, and the combined curve was used to estimate the velocity at day 50. 
Secondly, we applied the formula by \citet{nugent} 
$v_{ph} (50) ~=~ v_{ph}(t) \cdot (t / 50)^{0.464}$ for the last two published velocities
of SN~2005cs \citep{past}. The first method gave $v_{ph} \sim 2020$ km s$^{-1}$, while
in the second case the average of the predicted velocities was $\sim 2047$ km s$^{-1}$.
Finally, $v_{ph} (50) = 2030 \pm 300$ km s$^{-1}$ was adopted as an average.

Substituting these values into the formulae of \citet{ham1}, 
$D_V = 6.13 \pm 0.8$ Mpc and $D_I = 6.55 \pm 0.9$ Mpc was derived 
for the distance of SN~2005cs (Table \ref{const}).

Because SN~2005cs was a low-velocity SN, and such Type II-P SNe are represented
only by SN~1999br in the calibrating sample, we decided to recalibrate the
SCM including SN~2005cs as well. This new relation is expected to be better
constrained in the low velocity regime, thus, it may predict more accurate
distances for such SNe. 

Of course, one needs independent distances for such calibration. \citet{ham1}
used the Hubble-flow velocities for computing the distances. However, the host galaxy 
of SN~2005cs, M51, is too close to get reasonable estimates of its distance
from redshift.

\begin{figure}
\begin{center}
\psfig{file=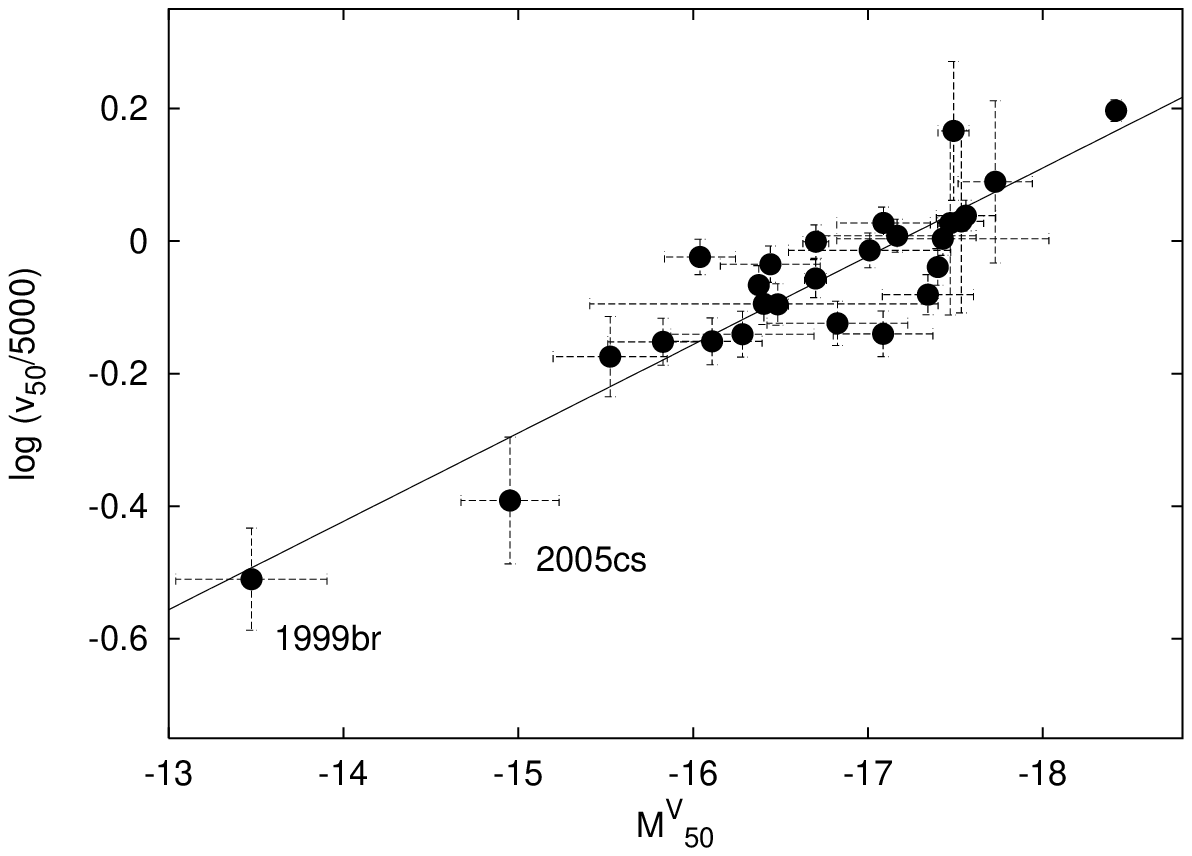,width=8cm}
\psfig{file=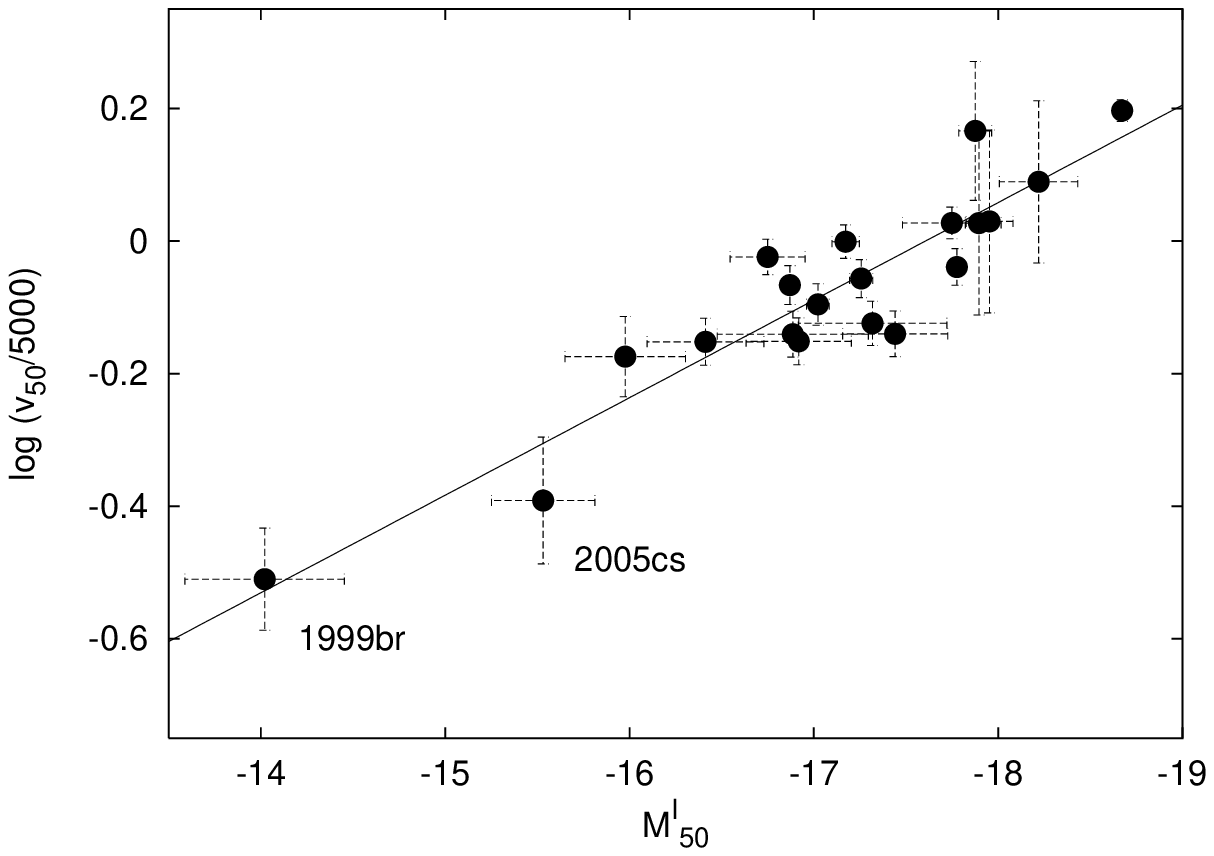,width=8cm}
\end{center}
\caption{The absolute magnitude -- velocity diagram in $V$ (top) and $I$ (bottom) 
 band. For SN~2005cs, the distance of $7.25$ Mpc was applied (see text). 
 The other 26 SNe in $V$ and 19 in $I$ are from \citet{ham1} and \citet{nugent}.}
\label{Mv}
\end{figure}

Therefore, we have adopted the weighted average of all the distances of
M51 except the value from SCM, using the reciprocal of
the errorbars as weights (see Table \ref{dist} in Section 3.1). This resulted
in $D = 7.25 \pm 1.21$ Mpc, which is slightly less than the average EPM-distance
($7.59 \pm 1.02$ Mpc) determined above. 
Fig. \ref{Mv} shows SN~2005cs on the absolute magnitude -- velocity diagram 
with the other calibrating SNe (26 in $V$- and 19 in $I$-band) from \citet{ham1} and
\citet{nugent}. It is seen that most of them have a velocity greater than $3000$ km/s. 
Only SN~1999br and SN~2005cs are in the low velocity regime. They have similar 
velocities, but SN~2005cs is brigther by almost $2$ magnitudes. 

The calibrating sample was fitted by the following equation of SCM:
\begin{equation}
 m - A + a \cdot log\left({v_{50} \over 5000}\right) = 5 \cdot log(H_0 D) - b
\label{3} 
\end{equation}
where $m$ is the observed magnitude (in $V$ or $I$), $A$ is the extinction, 
$H_0 ~=~ 73$ km (s Mpc)$^{-1}$ \citep{riess} is the Hubble-constant, 
$D$ is the distance, $a$ and $b$ are the fitted constants.

\begin{figure}
\begin{center}
\psfig{file=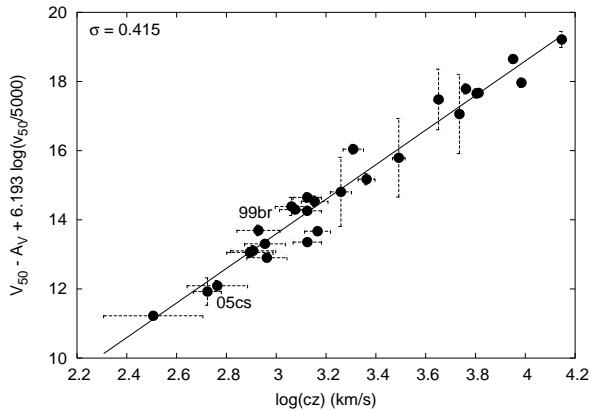,width=8cm}
\end{center}
\caption{The Hubble-diagram from $V$ magnitudes corrected for expansion 
 velocities. The constant $b$ (Eq. \ref{3}) is obtained from the fitted line.}
\label{scm_v}
\end{figure}

\begin{figure}
\begin{center}
\psfig{file=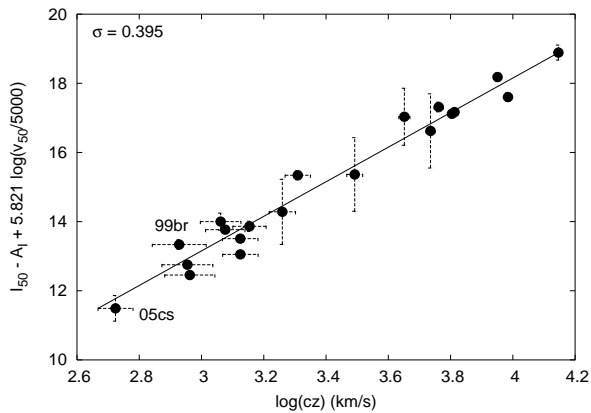,width=8cm}
\end{center}
\caption{The same as Fig. \ref{scm_v}, but for $I$ magnitudes.}
\label{scm_i}
\end{figure}

The results of the fit are seen in Fig. \ref{scm_v} ($V$) and in Fig. \ref{scm_i} ($I$). 
The fitted constants $a$ and $b$ are collected in Table \ref{const}. The inclusion of 
SN~2005cs changed mainly the constant $a$, the slope of the 
luminosity-velocity relation. 

With the new constants  the distance of SN~2005cs was re-evaluated. These 
values are also seen in Table \ref{const}. All of them are lower than the 
EPM-distance with fixed explosion epoch. The agreement is better with 
the EPM-distance with fitted $t_0$ (see the previous section). 
However, we stress that the present application of SCM is only preliminary,
because $i)$ it is based on velocities derived by extrapolation and $ii)$ 
the low-velocity SNe are represented very poorly in the calibrating sample
(only two objects including SN~2005cs itself), making the SCM-distance
more uncertain. We conclude that the SCM-distance to SN~2005cs is 
$D=6.36 \pm 1.30$ Mpc. 

\begin{table}
\begin{center}
\caption{
 The derived constants $a$ and $b$ of Eq. \ref{3} in $V$ and $I$ bands 
 and the same data from \citet{ham1}.  The obtained distances of 
 SN~2005cs (in Mpc) are also shown in the last column.}
\begin{tabular}{ccccc}
\hline
 & & a & b & D \\
\hline
 & \cite{ham1} & 6.564 (0.88) & 1.478 (0.11) & 6.13 (0.8) \\ 
 V & this paper & 6.193 (0.57) & 1.407 (0.08) & 6.35 (1.3) \\
\hline
 & \cite{ham1} & 5.869 (0.68) & 1.926 (0.09) & 6.55 (0.9) \\
 I & this paper & 5.821 (0.57) & 1.848 (0.09) & 6.37 (1.3) \\
\hline
\end{tabular}

\label{const}
\end{center}
\end{table}

\section{Discussion}

\subsection{The average distance to M51}

\begin{table}
\begin{center}
\caption{Comparison of the published and our distance estimates of M51. Errors are in
parentheses.}
\begin{tabular}{lcl}
\hline
 Method & Distance (Mpc) & Ref. \\
\hline
 YSA & 6.91 (0.67) & \citet{georg} \\
 PNLF & 7.62 (0.60) & \citet{ciard} \\
 SBF & 7.66 (1.01) & \citet{tonry} \\
 SN 1994I & 6.92 (1.02) & \citet{iwam} \\
 SN 1994I SEAM & 6.02 (1.92) & \citet{bar} \\
 EPM & 7.59 (1.02) & present paper \\
 SCM & 6.36 (1.30) & present paper \\
\hline
 average & 7.1 (1.2) & \\
\hline
\end{tabular}
\label{dist}
\end{center}
\end{table}

M51 is a very well-known galaxy, but its distance was determined only 
a few times between 1974 and 2005. \cite{sand} derived $9.6$ Mpc using sizes of H~II 
regions. Although they did not specify the errorbar of their result, it is probably
more uncertain than the other, more recent distance estimates.
\cite{georg} got $6.91 \pm 0.67$ Mpc from the photometric properties of 
young stellar associations (YSA). \cite{feld} determined $8.39 \pm 0.60$ Mpc using planetary 
nebulae luminosity function (PNLF), but later they revised it as $7.62 \pm 0.60$ Mpc using
improved reddening \citep{ciard}. From surface brightness fluctuation (SBF) 
\cite{tonry} obtained $7.66 \pm 1.01$ Mpc. 

The Type Ic SN~1994I also gave a chance for distance determination to M51. By 
fitting theoretical light curves to observations \cite{iwam} got $D= 6.92 \pm 1.02$ 
Mpc. With the spectral-fitting expanding atmosphere method (SEAM) \cite{bar} 
derived $6.02 \pm 1.92$ Mpc.

Table \ref{dist} contains the various M51 distances (except that of \citet{sand}, 
which deviates mostly from all the other ones) together with our results based on SN~2005cs. 
Adopting their average, the final distance to M51 turns out to be
$$D_{M51} ~=~ 7.1 ~\pm 1.2 ~\rm{Mpc}.$$

\subsection{Physical properties of SN~2005cs}

\begin{table*}
\begin{center}
\caption{The inferred physical parameters of SN 2005cs, based on $D=7.1$ Mpc and
$E(B-V) = 0.11$ mag. See text for explanation. The errors are in parentheses.}
\begin{tabular}{ccccccccc}
\hline
$M_V(50)$ & $M_I(50)$ &$M_I^{prog}$ & $v_{ph}(50)$ & $M_{Ni}$ & $E_{expl}$ & $M_{ej}$ & $R_{ini}$ & $M_{prog}$\\
(mag) & (mag) &(mag) &(km s$^{-1}$) & ($M_\odot$)& ($10^{51}$ erg) & ($M_\odot$) & ($R_\odot$) & ($M_\odot$) \\
\hline
-14.88 & -15.46 & -5.5 & 2030 & 0.009 & 0.19 & 8.3 & 177 & 9.6 \\
(0.3) & (0.3) & (0.7) & (300) & (0.003) & (0.15) & (5.3) & (150) & (5.3)\\
\hline
\end{tabular}
\label{pars}
\end{center}
\end{table*}

The physical parameters of SN~2005cs are sensitive to the distance used
to derive these parameters. All the previous studies \citep{li,maund,past,tsvet}
adopted the SBF-distance $D = 8.39$ Mpc by \citet{feld}, which was actually shortened
to $7.62$ Mpc by the same authors in a subsequent paper \citep{ciard}.   
Thus, the published parameters of SN~2005cs may need some revision based on the
updated M51 distance.

We have calculated some of the physical parameters adopting the average M51 distance
($D = 7.1$ Mpc) from the previous section. Using the light curves of \citet{tsvet},
the middle-plateau absolute magnitudes (corrected for reddening with $E(B-V)=0.11$)
are $M_V(50) = -14.88 \pm 0.3$ and $M_I(50) =-15.46 \pm 0.3$. 
For comparison, \citet{tsvet} determined
$M_V(50) = -15.33$, which is $\sim 0.4$ mag brighter, because of their longer adopted distance. 
From the correlation between the Ni-mass and $M_V$ \citep{ham0}, the calculated
Ni-mass is $M_{Ni} \approx 0.009 \pm 0.003 ~M_\odot$. Again, this is somewhat less
than the value given by \citet{tsvet} ($\sim 0.018 ~M_\odot$). However, this less amount
of synthesized Ni is in better agreement with the relation of \citet{ham0} between
the Ni-mass and the middle-plateau photospheric velocity. For $M_{Ni} \sim 0.009 ~M_\odot$ 
this relation predicts $v_{ph}(50) \approx 2200$ km s$^{-1}$, which is in good agreement
with $\sim 2030$ km s$^{-1}$ determined in Sect. 2.2. \citet{tsvet}
estimated $v_{ph}(50) \sim 2600$ km s$^{-1}$, which is too high, since the observed velocity
is already $\sim 2200$ km s$^{-1}$ at day +35 (see Table \ref{data}).

The physical parameters for the progenitor star have been derived from the 
formulae given by \citet{nady}.
These equations relate the explosion energy, ejected envelope mass and initial radius to 
the plateau absolute magnitude, explosion velocity and plateau duration. Using
$M_V \sim -14.88$ mag, $v_{ph} \sim 2030$ km s$^{-1}$ estimated above, 
and $\Delta t_p \sim 86$ days from the light curves of \citet{tsvet}, 
$E_{expl} = 0.19^{+0.17}_{-0.10} \cdot 10^{51}$ erg, $M_{ej} = 8.3^{+6.8}_{-4.0} ~M_\odot$ and 
$R_{ini} = 177^{+258}_{-100} ~R_\odot$ has been calculated for the explosion energy, envelope
mass and radius, respectively. These are in good agreement with the ones
given by \citet{tsvet}, despite the somewhat shorter distance applied in this
paper. The absolute magnitude of the progenitor was also updated using the magnitude
estimates from $HST$-photometry. For the observed brightness of the likely progenitor star, 
\citet{li} reported $I = 24.15 \pm 0.2$ mag, while \citet{maund} got 
$I \approx 23.3$ mag, about 1 mag brighter. The average of these is $\sim 23.7 \pm 0.6$ 
mag, which is adopted here. Using the $7.1$ Mpc distance, the absolute $I$-band 
magnitude is $M_I^{prog} \approx - 5.5 \pm 0.7$ mag, which is the same as the result of
\citet{li}. Although their progenitor brightness is fainter than the average value
used here, they adopted a longer distance to M51, leading to the same absolute megnitude.
 
The inferred physical parameters of SN~2005cs and its progenitor 
are summarized in Table \ref{pars}. 

From the envelope mass of $\sim 8 ~M_\odot$, the progenitor mass $M_{prog}$
can be calculated by adding the estimated mass of the compact remnant of
the core-collapse process. Assuming that it is a neutron star, its mass
is estimated as $\sim 1.3 \pm 0.5 ~M_\odot$ \citep{fryer}. The progenitor
mass is then $M_{prog} = 9.6 \pm 5.3 ~M_\odot$, which is in 
good agreement with the mass of $7$ - $9 ~M_\odot$ estimated
from the direct detection of the progenitor \citep{li, maund}, as it was
also found by \citet{tsvet}. On the other hand, it seems to be in
contrast with the theoretical prediction by \citet{zamp} that 
low-luminosity SNe II-P, such as SN~1999br and SN~1997D
may occur from a peculiar, low-energy explosion of a more massive supergiant star
of $\sim 15$ - $20 ~M_\odot$. SN~2005cs was definitely such a low-energy, 
low-luminosity SN, as it is also indicated by its low expansion velocity
\citep{past}, because the explosion energy derived above is the lowest
among the energies of other SNe II-P determined in similar way (see Table~2
of \citet{nady}). It is even lower than the one inferred by \citet{zamp}
for SN~1999br ($\sim 0.6 \cdot 10^{51}$ erg), which had lower expansion
velocity than SN~2005cs. 
Although \citet{past} argued that the mass of
the progenitor may be underestimated from its observed $I-$band flux, the
explosion characteristics of SN~2005cs are in better agreement with the
low-mass progenitor scenario proposed by \citet{li} and \citet{maund}.

The consistency of the inferred progenitor mass and radius can be tested with 
the prediction of the evolutionary tracks. We have applied the Padova
evolutionary models \citep{bressan} with $Z = 0.02$ and $Z = 0.008$.
For both metallicities, the $M = 7 ~M_\odot$ models have $R \sim 180 ~R_\odot$ 
radius at K3 - K4 spectral type ($T_{eff} \sim 4000$ K), 
near the end of their calculated evolutionary track. This is also in good
agreement with the K - M spectral type estimated from the $HST$ flux limits
\citep{li, maund}. The $M > 9 ~M_\odot$ models have $R > 300 ~R_\odot$
in this spectral type regime, regardless of metallicity. Note, however, that
the progenitor parameters estimated above are quite uncertain, therefore
the larger, more massive progenitor scenario cannot be ruled out from these
data alone. It is concluded that the available information may suggest
a consistent picture for SN~2005cs and its progenitor, namely a low-mass, 
($M \sim 9 ~M_\odot$) K3 - K4 spectral type supergiant that showed a low-energy
explosion ($E_{expl} \sim 0.2 \cdot 10^{51}$ erg) producing an underluminous
($M_V(50) \sim -15$ mag), slowly expanding ($v_{ph}(50) \sim 2000$ km s$^{-1}$)
Type II-Plateau SN. 
 
\section*{Acknowledgments}
The authors are grateful to Andrea Pastorello for sending the velocity
data of SN~2005cs in digital form, and to the referee, David Branch for
the valuable comments and suggestions that improved the paper. 
This work has been supported by Hungarian OTKA Grants TS~049872 and T~042509.
The NASA Astrophysics Data System, the SIMBAD and NED databases, 
the Canadian Astronomy Data Centre and the Supernova
Spectrum Archive (SUSPECT) were used to access data and references. 
The availability of these services are gratefully acknowledged.

\label{lastpage}

\end{document}